\documentclass[pdflatex,sn-nature]{sn-jnl}

\usepackage{graphicx}%
\usepackage{multirow}%
\usepackage{amsmath,amssymb,amsfonts}%
\usepackage{amsthm}%
\usepackage{mathrsfs}%
\usepackage[title]{appendix}%
\usepackage{xcolor}%
\usepackage{textcomp}%
\usepackage{manyfoot}%
\usepackage{booktabs}%
\usepackage{algorithm}%
\usepackage{algorithmicx}%
\usepackage{algpseudocode}%
\usepackage{listings}%
\usepackage{subcaption}
\usepackage{float}
\usepackage{fvextra}
\usepackage{booktabs}
\usepackage{placeins}

\theoremstyle{thmstyleone}%
\theoremstyle{thmstyletwo}%

\theoremstyle{thmstylethree}%

\begin{document}

\title[Article Title]{
LLM Facilitation of Group Decision-Making

}


\author*[1]{\fnm{José} \sur{Córdova}}\email{josecordova2027@u.northwestern.edu}
\author[2]{\fnm{Chun-Wei} \sur{Chiang}}
\author[3]{\fnm{Miriam} \sur{Schirmer}}
\author[2]{\fnm{Ming} \sur{Yin}}
\author*[1]{Em\H{o}ke{-}\'{A}gnes Horv\'{a}t}\email{a-horvat@northwestern.edu}

\affil[1]{\orgname{Northwestern University}, \city{Evanston}, \state{IL}, \country{United States}}
\affil[2]{\orgname{Purdue University}, \city{West Lafayette}, \state{IN}, \country{United States}}
\affil[3]{\orgname{University of Konstanz}, \city{Konstanz},  \country{Germany}}


\abstract{Human groups often fail to integrate information distributed across their members, converging on incorrect decisions. We ask whether large language models (LLMs) can help groups reach the correct decision in such settings. In an online, chat-based experiment, 138 three-person groups (N = 414) completed a hidden-profile hiring task that required selecting the best candidate using public information available to all group members and private information available only to individual members. The randomized human-subject experiment included three conditions: no facilitation, an individual-level LLM that messaged each member privately, and a group-level LLM that messaged the whole group. The LLM facilitation consisted of adaptive interventions that summarized the discussion, nudged quiet members, and played devil’s advocate. The LLM did not know which candidate was correct. Both LLM conditions significantly increased the rate at which groups reached consensus on the correct candidate, from 32.6\% without facilitation to 56.5\% with LLM facilitation. LLM-facilitated groups exchanged more consensus messages signaling agreement on the group’s choice. Participants in the group-level condition also disclosed significantly more information about the correct candidate, whereas those receiving individual-level facilitation repeated public information less often. Finally, group-level facilitation received more favorable usability ratings than individual-level facilitation. These results show that LLMs can deliver adaptive, theory-informed interventions that improve group decision-making, and that the design of these interventions may shape how people respond to them.}


\maketitle

\section*{Main}\label{sec_main}

Many of the most consequential decisions are made by groups. Emergency medical teams, jury rooms, hiring committees, and scientific collaborations all rest on the premise that a group of people will reach a better judgment than any one of them could alone \cite{Woolley2010-pk_evidence,Surowiecki2005-cw}. In principle, groups can outperform individuals by pooling diverse perspectives and knowledge \cite{Page2007-ci}. Yet, collective decision-making often falls short of this ideal, as a range of factors undermine group performance. Groups often struggle to integrate information, let a few voices dominate deliberation, suffer from information overload, or settle into premature consensus \cite{Kameda2022_information_ci, Kerr2004-gw,Centola2022_CI,Lu2011_hidden_survey,Brodbeck2007_distr_knowledge}. The costs of these failures can be catastrophic. For example, the 9/11 Commission investigating the attacks found that different agencies each held clues to the plot, but a culture that restricted information sharing prevented any agency from assembling the full picture. As the Commission concluded, ``The biggest impediment to all-source analysis—to a greater likelihood of connecting the dots—is the human or systemic resistance to sharing information'' \cite{Kean2004_911report}. It is thus imperative to study how groups can be supported in reaching better collective judgments. 

In this paper, we ask whether LLM facilitation can help groups reach more accurate decisions. Recent advances in LLMs open new avenues for supporting critical decision-making in groups \cite{Woolley2025-dz_opt_ci,Gonzalez2026_complementarity,Cui2024-dm-enhancedci,Burton2024-le-reshapeCi}. LLMs could facilitate decision-making not by supplying the right answer, but by tracking a human conversation and intervening on the \emph{process} by which the group decides. LLMs can potentially improve collective reasoning, memory, and attention \cite{Cui2024-dm-enhancedci,Burton2024-le-reshapeCi, Woolley2025-dz_opt_ci}. However, groups may over-rely on the assistant \cite{butler2025nfw,overreliance_raees,Vaccaro2024-hn}, disengage from the decision \cite{Gonzalez2026_complementarity,Fgener2021_borgs}, or converge too easily, eroding the diversity that makes a group valuable \cite{Burton2024-le-reshapeCi,Doshi2024_creativity_ind_Vs_col,Anderson2024_homogeneization_ideation}. The challenge then is to design LLM facilitation that leaves the judgment to humans, supports their collective reasoning process, and avoids common collective decision-making pitfalls.

To develop and test LLM facilitation in a controlled setting, we use the hidden profile task, in which a group must integrate information distributed across its members to reach the correct decision \cite{Stasser1992}. In this task, a group must choose the best of several options. Each member receives only a partial set of information, and the group can identify the correct option only if members share the information they hold privately \cite{MesmerMagnus2009_infosharing, SchulzHardt2012_how_synergy}. In practice, groups discuss information everyone already has more often than information held by a single member \cite{Lu2011_hidden_survey,Sohrab2015_HP_rev}, partly for three well-documented reasons. Information overload makes it hard to track many attributes of the options at once \cite{Stasser1987-st}. Unequal participation leaves quieter members' knowledge unrevealed \cite{Kerr2004-gw,Woolley2010-pk_evidence}. And premature convergence locks a group onto an early favorite before its members weigh the alternatives \cite{mojzisch2010knowing}.

To target each of these challenges, we designed three LLM interventions drawing on techniques previously shown to improve group processes. To ease information overload, the LLM facilitator \textit{summarized} the discussion \cite{Alsobay2025-al}. To address unequal participation, it \textit{nudged} members who had gone quiet or contributed little \cite{gupta_nudging}. To counter premature convergence, it offered a \textit{devil's advocate} message presenting an alternative option when a consensus appeared to be emerging \cite{Chiang2024-ly}. In all cases, the LLM facilitator was blind to the candidates' attribute values, so it intervened on the discussion process without knowing which choice was correct. This design leverages the capabilities of LLMs to make interventions more adaptive to the conversation in two ways. First, it improves the timing of interventions. Rather than intervening on a fixed schedule or relying on simple conversational heuristics, the LLM facilitator tracked the conversation and assessed whether an intervention would help. Second, it tailors the text of each intervention to the content of previous messages.

We deployed these three interventions in two practically relevant scenarios. First, individuals are increasingly bringing LLM assistants into meetings. Second, organizations are beginning to integrate shared assistants into collaborative workspaces, such as Claude Tag \cite{claude_tag}. To capture these emerging use cases in a controlled setting, we embedded LLM facilitation in group decision-making at either the individual or the group level. The first approach employed an LLM facilitator as a private aid to \textit{each member}, delivering interventions that only one person sees. The second approach used a shared facilitator, delivering the same three types of interventions to the group as a whole. This distinction may matter in practice, since private versus public facilitation messages can change who speaks up and how people respond to the facilitator. Previous research has compared these communication strategies, finding that nudging people privately is more effective than nudging them publicly \cite{Do2022_groupvsprivate}. Here, we extend this comparison by leveraging LLM capabilities to adapt the timing, type, and text of interventions to the needs of the group or individual, rather than fixing the number and type of interventions across conditions. We experimentally tested these two approaches with 138 three-person groups ($N=414$) completing a hidden profile hiring task, randomly assigning groups to individual-level LLM facilitation, group-level LLM facilitation, or no facilitation. We chose no facilitation as our baseline because most group interactions take place without any facilitator, and these are the settings where scalable technology could have the largest impact. Our research focus is therefore not on whether LLMs can replace a human facilitator, but on whether LLM facilitation can help in settings where human facilitation is often infeasible because of its high cost and logistical difficulty.

Our work contributes to the human-AI interaction literature, specifically on collective decision-making, by showing that adaptive summaries, nudges, and devil's advocate interventions grounded in group decision-making theory can improve accuracy, regardless of whether the LLM addresses the group collectively or each member privately. 

\section*{Results}\label{sec_results}

\subsection*{Experimental design}

We built an online chat platform for groups of three participants to select the best of three faculty candidates in a hidden profile task setup, in line with specifications used in prior literature \cite{Stasser1985-rd,Lu2011_hidden_survey, hidden_profile_faculty}. Each participant saw six attributes per candidate, rated as average, good, or exceptional. Four attributes were public (available to everyone in the group) and two were private to each member, so the group as a whole held ten attributes per candidate (see Table \ref{sitab:attributes} for the complete list). Participants were instructed that the correct candidate was the one with the most exceptional attributes. By design, there was one correct candidate. However, each participant’s information pointed to a different candidate, so identifying the correct candidate required participants to disclose and combine their private information. Groups discussed the decision in a text chat for a minimum of five minutes, then each member submitted a final choice, knowing that they would receive a bonus if all group members unanimously voted for the best candidate. Figure \ref{sifig:chatbox} shows examples of the chat interface.

\begin{figure}[h]
    \centering

    \includegraphics[width=0.98\textwidth]{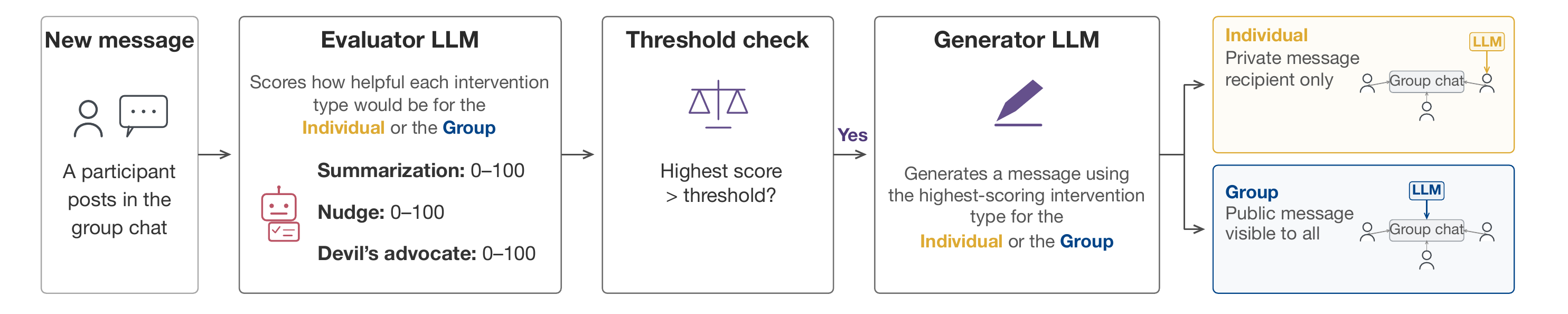}
     \caption{\textbf{Two-stage system determining intervention delivery in LLM-facilitated groups.} The system checks the group chat every 10 seconds for new participant messages. When a new message is detected, the evaluator LLM (GPT-5) scores three possible interventions (a summary, a nudge, or a devil’s-advocate response) from 0 to 100 based on their expected helpfulness, either to the individual it assists or to the group as a whole. Text \ref{si:prompts_evaluator} shows the prompts used for the evaluator LLM. If the highest score is higher than the intervention threshold, the generator LLM (GPT-5) produces a message for the corresponding intervention type using the conversation as context (see prompts for the generator LLM in Text \ref{si:prompts_generator}). Before delivery, the system checks for messages that arrived during evaluation or generation; if a new message has arrived, the intervention is not sent, and the process starts again using the latest conversation. Otherwise, the intervention is delivered privately in the individual condition or publicly in the group condition.}
    \label{fig:llm_logic}
\end{figure}

Figure \ref{fig:llm_logic} shows the logic that determined when the LLM facilitator intervened. Each time a participant sent a message, an \textit{evaluator LLM} guided by a structured prompt assigned a score between 0 and 100 to each of the three interventions, based on specific conversational characteristics. For example, the summarization score was higher when multiple attributes were discussed in parallel or when participants asked about what attributes had been covered. The nudging score was higher when a participant had not contributed at all, or when several messages from the rest of the group have occurred without a substantive message from that participant. Finally, the devil's advocate score was higher when participants agreed rapidly without exploring alternatives and no objections or critical questions were raised. 

The intervention with the highest score (the one the evaluator LLM deemed most appropriate) was checked against a threshold. We used a scoring guide in our prompt to set the threshold (see Text \ref{si:prompts_evaluator}) at 70 out of 100, which corresponds to an above-moderate score according to the guide’s anchor values. We also tested pilot conversations to ensure that the LLM facilitator neither intervened excessively (e.g., interrupting productive exchanges or repeating the same type of intervention in succession) nor insufficiently (e.g., failing to nudge a silent participant). If the score was less than or equal to the threshold, the conversation continued without an intervention. If it was higher, a \textit{generator LLM} created the intervention message for the selected intervention type, allowing the facilitator to respond to the conversational context. The message was then delivered either privately to a single participant by their own facilitator (individual-level LLM) or to the entire group by the shared facilitator (group-level LLM), depending on the experimental condition. This adaptive intervention timing is an important feature of our system, because previous research on LLM cues in brainstorming found that participants experienced badly timed interventions as disruptive \cite{cueing_the_crowd_dow}.

We recruited 414 participants via Prolific and randomly assigned them to 138 groups, which were in turn randomly assigned to one of three conditions: no facilitation, individual-level LLM, or group-level LLM. Our primary outcome was \textit{decision accuracy}, measured as the percentage of groups in which all three members chose the correct candidate. We manually coded the $3{,}655$ human messages to establish which candidate and which specific attribute each message discussed, which allowed us to compare conditions in how often participants discussed public versus private attributes. We also coded messages for their communicative function, assessing their role in the conversation beyond their informational content \cite{bales1950interaction}. This approach has been used extensively in the analysis of group conversations \cite{gouran1996functional,kolbe2011microco}, and we adapted previously used categories to our task. We report Cohen’s kappa ($\kappa$) from two independent human coders on 22\% of the transcripts. Each human message was coded as one of the following: eliciting information (e.g., ``What are y'all's exceptionals for A?''; $\kappa = 0.78$), contributing specific information (``publications is exceptional for candidate B’’; $\kappa = 0.79$), contributing non-specific information (``candidate A has many exceptional qualities'' without specifying which ones; $\kappa = 0.76$), coordinating the decision by managing how the group would choose (``so who shall we choose?’’; $\kappa = 0.71$), or reaching consensus by agreeing with a candidate choice (``I can agree with A too’’; $\kappa = 0.66$). Text \ref{si:codebook} shows the complete codebook. Finally, all participants completed a cognitive load survey (5-item NASA-TLX) \cite{HART1988139_nasatlx}. Participants in the two LLM conditions additionally completed an LLM usability survey, in which they rated eight items on 5-point Likert scales and answered one optional open-ended question (see full survey items in Text \ref{si:sec_survey_questions}). 

Regardless of experimental condition, groups exchanged a similar number of messages, with similar response latency and similar inequality in the number of words contributed by each participant (see Figure \ref{sifig:con_characteristics}A--C). In terms of intervention delivery, nearly all participants were exposed to a summary, and most also experienced a devil's advocate intervention. Roughly one in four participants in the individual condition and three in five in the group condition were nudged during the discussion (Figure \ref{sifig:con_characteristics}D). This variation in how the interventions reached the two facilitated conditions is consistent with the public-versus-private contrast that defines the manipulation. With this groundwork in place, we turn to whether LLM facilitation changed decision accuracy.

\subsection*{LLM facilitation raises decision accuracy}

We found that groups with LLM facilitation reached the correct decision more often than groups without facilitation, both when the two LLM conditions were analyzed separately and when they were pooled. The individual-level and group-level LLM conditions each yielded 26 correct decisions out of 46 groups, so they performed equally in our sample (Table~\ref{tab:accuracy}). Each LLM condition selected the correct candidate significantly more often (56.5\%) than groups deciding on their own (32.6\%), corresponding to an odds ratio of 2.69 (95\% CI $[1.15, 6.28]$, $p = 0.022$) from a logistic regression with the no facilitation condition as the reference category. When we pooled the two LLM conditions, the effect was medium-sized (Cohen’s $h = 0.49$), with a narrower confidence interval on the odds ratio (95\% CI $[1.28, 5.64]$) and stronger evidence of a difference between LLM facilitation and no facilitation ($p = 0.009$). See Table~\ref{tab:app_accuracy_logit} for the full regression output.

\begin{table}[h]
\caption{Decision accuracy by condition. Both individual and group-level LLM facilitation yielded significantly higher accuracy than groups without facilitation. The two LLM conditions did not differ from each other in decision accuracy.}\label{tab:accuracy}
\begin{tabular}{lcccc}
\toprule
Condition & \# Correct Groups & Accuracy \% (95\% CI) & OR (95\% CI) & $p$ \\
\midrule
No facilitation        & 15 & 32.6 (20.9--47.0) & reference              & --   \\
Individual-level LLM & 26 & 56.5 (42.2--69.8) & 2.69 (1.15--6.28) & 0.022 \\
Group-level LLM      & 26 & 56.5 (42.2--69.8) & 2.69 (1.15--6.28) & 0.022 \\
\botrule
\end{tabular}

\end{table}
Having established the effect of LLM facilitation on decision accuracy, we examined how it shaped deliberation. We first analyzed information sharing, measuring how often groups repeated public attributes and how many private attributes they discussed. We then coded each message by communicative function and compared the distribution of functions across conditions.

\subsection*{LLM facilitation shapes information sharing}
\label{sec_repetition}

\begin{figure}[htbp]
    \centering

    \begin{minipage}[t]{0.49\textwidth}
        \centering
        \includegraphics[width=\linewidth]{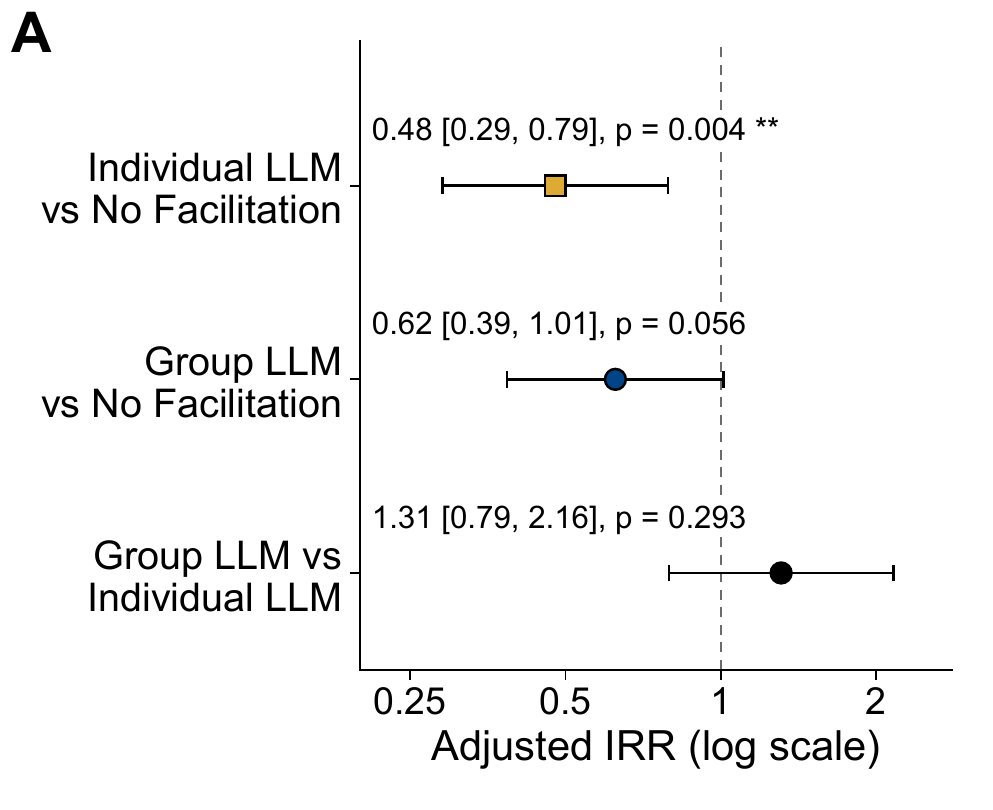} 
    \end{minipage}
    \hfill
    \begin{minipage}[t]{0.49\textwidth}
        \centering
        \includegraphics[width=\linewidth]{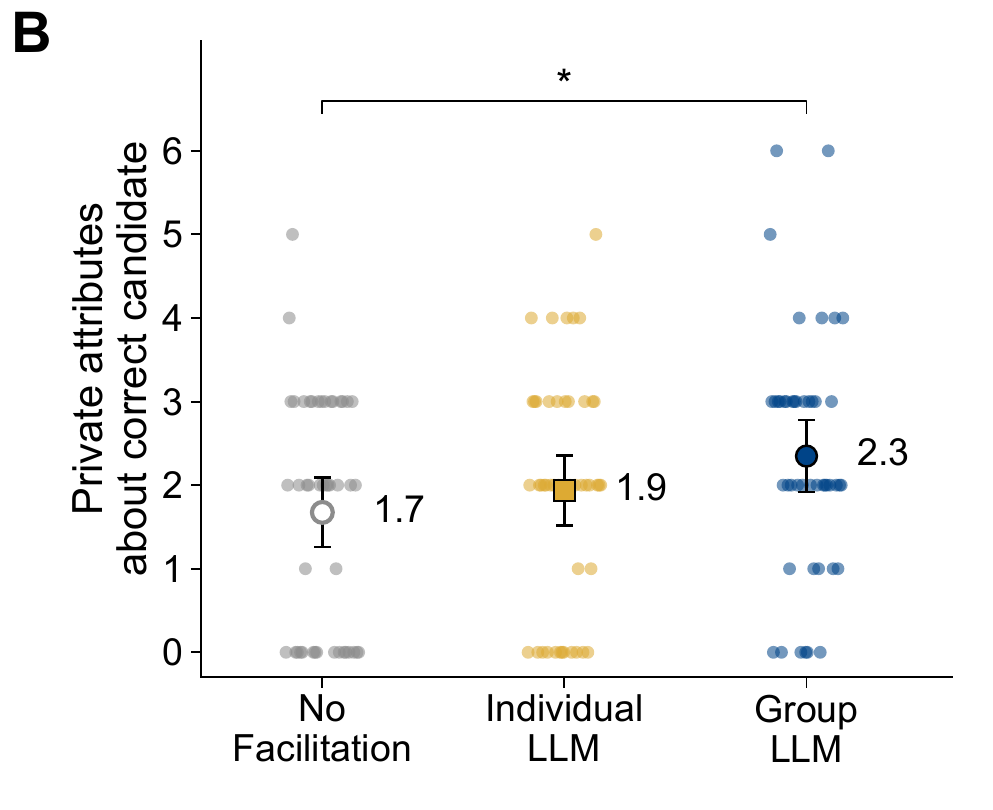} 
    \end{minipage}

    \caption{Repetition of public attributes and disclosure of private information, by condition. \textbf{A}: Adjusted incidence rate ratios (IRRs) for repetition of public attributes during group discussion. IRRs are reported from a negative binomial regression that controls for the number of distinct attributes discussed and the number of messages sent. Error bars show 95\% CIs and the dashed line marks $\text{IRR} = 1$. \textbf{B}: Number of private attributes about the correct candidate shared by groups. Asterisks mark significant differences: $^{*}p < 0.05$;  $^{**}p < 0.01$. Participants in the individual-level condition repeated public attributes less often than groups without facilitation. Participants in the group-level LLM facilitation condition revealed more private attributes than groups without facilitation.}
    \label{fig:information_disclosure_and_repetition}
\end{figure}

Perhaps the most important finding in the literature on the hidden profile paradigm is the common-knowledge effect: groups tend to focus their discussion on information held by everyone \cite{Gigone1993_common_knkowledge,Stasser1985-rd}. We therefore examined how often groups \textit{repeated public information}, operationalized as mentions of a public attribute beyond its first occurrence in the group discussion. Raw counts did not differ significantly across conditions (Figure \ref{sifig:information_disclousre_and_repetition_appendix}A), in part because repetition is confounded with discussion volume: groups that surface more attributes or send more messages have more opportunities to repeat them. To isolate the condition effect, we ran a negative binomial regression of repetition counts on condition, controlling for the number of distinct attributes a group discussed and the number of messages it sent. The resulting coefficients are shown in Figure \ref{fig:information_disclosure_and_repetition}A (full output in Table \ref{tab:app_repetition_nb}). Groups in the individual-level LLM condition repeated public attributes significantly less often than groups without facilitation ($\text{IRR} = 0.48$, $p = 0.004$), which corresponds to roughly half as many repetitions. The group-level LLM condition did not differ significantly from no facilitation ($\text{IRR} = 0.62$, $p = 0.056$), and the two LLM conditions did not differ significantly from each other ($p = 0.293$). In other words, for a given number of distinct attributes discussed and messages sent, groups in the individual-level condition devoted less of their discussion to restating public information.

Reducing the repetition of public attributes can be helpful, but it is not enough on its own to reach the correct decision. In the hidden profile task, \textit{disclosure of private information} is particularly important for accurate decision-making. Figure \ref{fig:information_disclosure_and_repetition}B shows the number of private attributes about the correct candidate that were revealed during group discussion, that is, attributes initially held by only one group member. Consistent with our preregistered directional hypothesis (AsPredicted $\#280,167$) and with prior work suggesting greater information sharing in groups with LLM facilitation \cite{Alsobay2025-al}, we conducted one-tailed Welch’s $t$-tests comparing each facilitated condition to groups without facilitation. Groups without facilitation disclosed an average of 1.7 private attributes about the correct candidate, compared with 1.9 in the individual-level LLM condition and 2.3 in the group-level LLM condition. The difference between the no facilitation and group-level LLM conditions was significant ($p = 0.014$), whereas the difference between the no facilitation and individual-level LLM conditions was not ($p = 0.188$). The two LLM conditions also did not differ significantly from each other ($p = 0.084$). The same pattern held for the total number of attributes discussed about the correct candidate, including both private and public attributes (see Figure \ref{sifig:information_disclousre_and_repetition_appendix}B). We return to the non-significant result for the individual-level LLM condition in the \textit{Discussion}.

\subsection*{Facilitation increases consensus messages}

\begin{figure}[htbp]
    \centering
    \includegraphics[width=0.8\textwidth]{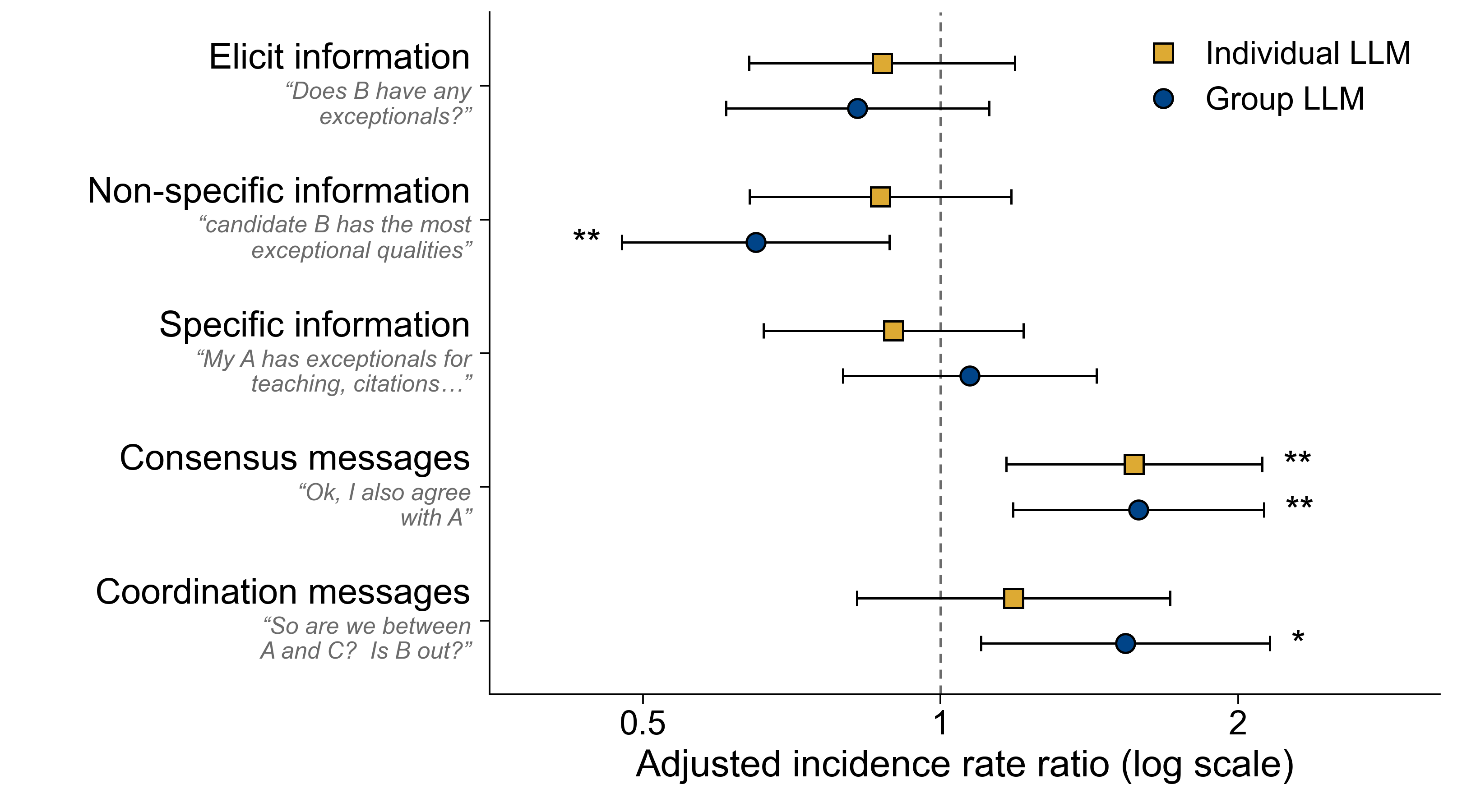}
    \caption{Incidence rate ratios (IRRs) for each coded communicative function, comparing the individual (yellow) and group (blue) LLM conditions to no facilitation. Estimates come from negative binomial regressions of per-group message counts for each communicative function, with the total number of messages a group sent included as a covariate. Error bars are 95\% confidence intervals on a log scale. Asterisks mark significant differences: $^{*}p < 0.05$;  $^{**}p < 0.01$. Both individual and group-level LLM facilitation yielded more consensus messages than no facilitation. The group-level LLM condition also yielded more coordination messages and fewer nonspecific-information messages.}
    \label{fig:language_type}
\end{figure}

In addition to information-sharing behavior, we analyzed the communication between group members to understand how groups solicited and integrated information to reach a decision. This matters because groups can perform poorly even when all the information is available \cite{SchulzHardt2012_how_synergy}. Previous research has analyzed group members’ behavior to understand how groups coordinate their discussion and arrive at decisions \cite{gouran1996functional,bales1950interaction,kolbe2011microco}. Following this literature, we coded the types of messages relevant to our setting. For each group, we counted how often members elicited information, contributed specific or nonspecific information, coordinated the decision, and built consensus by agreeing with a candidate choice. For each of these communicative functions, we fitted a negative binomial regression of the per-group count on condition, including the number of messages a group sent as a covariate to account for differences in how much groups talked. Figure~\ref{fig:language_type} reports the resulting IRRs relative to no facilitation.

The strongest effects were on consensus messages: both forms of LLM facilitation produced roughly 60\% more of them than no facilitation (group-level LLM: $+59\%$, 95\% CI $[+18\%, +112\%]$, $p = 0.002$; individual-level LLM: $+57\%$, 95\% CI $[+16\%, +111\%]$, $p = 0.003$). This is consistent with the groups’ voting behavior: both LLM conditions were more likely than groups without facilitation to reach complete consensus in their votes (individual-level LLM: $\text{OR} = 4.11$, 95\% CI $[1.62, 10.42]$, $p = 0.003$; group-level LLM: $\text{OR} = 4.75$, 95\% CI $[1.82, 12.36]$, $p = 0.001$). Group-level LLM facilitation additionally shaped two other behaviors: it increased coordination messages by 54\% (95\% CI $[+10\%, +115\%]$, $p = 0.012$) and reduced nonspecific contributions by 35\% (95\% CI $[-52\%, -11\%]$, $p = 0.007$). Full output is shown in Table~\ref{tab:app_speech_acts_nb}. These results describe how groups talked, but how did participants themselves rate their experience with the LLM facilitator?

\subsection*{Usability measures favor group-level LLM facilitation}

An LLM facilitator that improves group decision-making becomes practically useful only if people are willing to work with it. Understanding users’ experience is therefore key, because it predicts whether people are likely to adopt the facilitator and keep using it \cite{Venkatesh2003_survey_use,Glikson2020_trustinai}. To understand this experience, participants in the two LLM conditions rated the facilitator on eight usability items, such as its perceived quality, how satisfying it was to work with, and how much it felt like a collaborator. See Text~\ref{si:sec_survey_questions} for the full survey items.

\begin{figure}[htbp]
    \centering
    \includegraphics[width=1\textwidth]{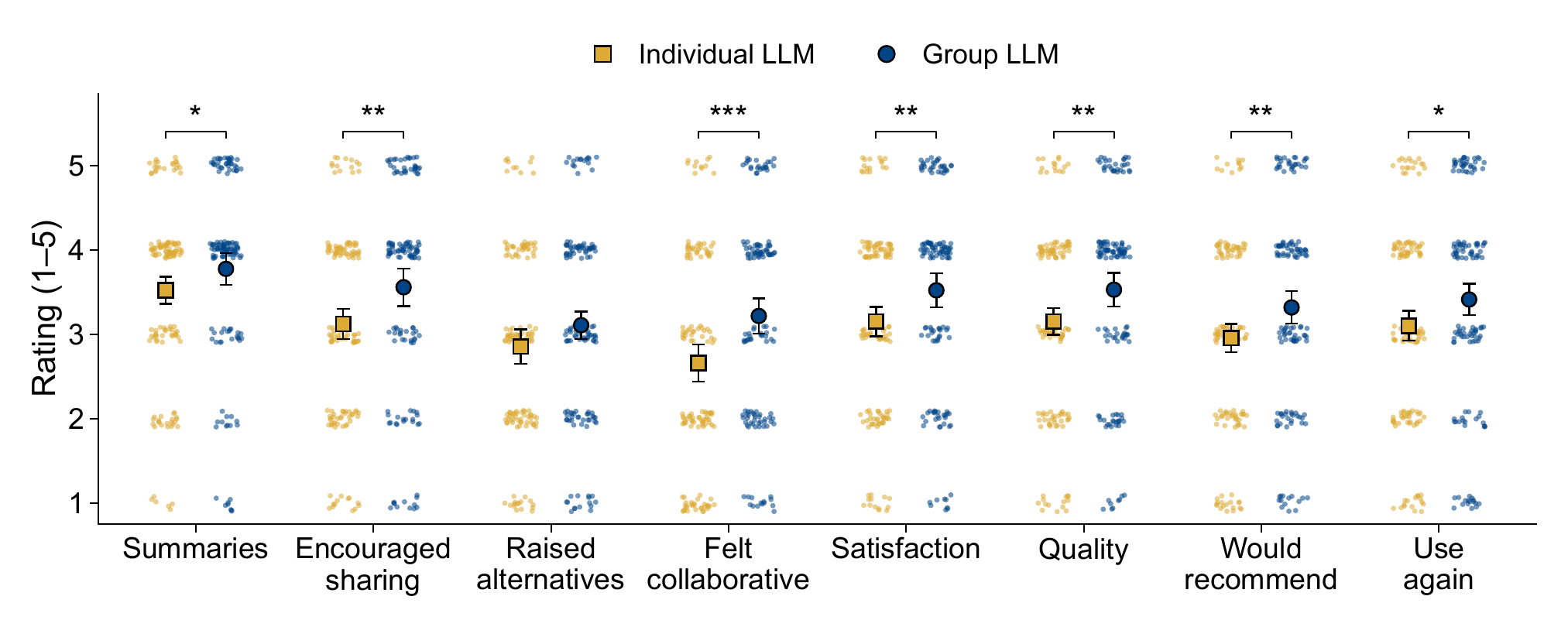}
    \caption{Post-task ratings of the LLM facilitator on eight usability items, by condition. Dots are condition means with 95\% confidence intervals; faded, jittered dots are individual participant responses. Asterisks mark significant differences from per-item OLS regressions with standard errors clustered by group: $^{*}p < 0.05$;  $^{**}p < 0.01$; $^{***}p < 0.001$. Group-level LLM facilitation was rated more favorably than individual-level LLM facilitation on seven of eight items.}
    \label{fig:llm_usability}
\end{figure}

Group-level LLM participants rated the facilitator more favorably than individual-level LLM participants on every item (Figure~\ref{fig:llm_usability}). Seven of the eight differences were significant, with small-to-moderate effect sizes. The largest was for the feeling of collaborating with the facilitator (Cohen’s $d = 0.44$, $p < 0.001$), followed by encouraging information sharing ($d = 0.36$, $p = 0.003$), the quality of the LLM at facilitating the task ($d = 0.32$, $p = 0.004$), and satisfaction with the facilitation the LLM provided ($d = 0.31$, $p = 0.007$). The only item that missed significance was whether the facilitator raised useful alternatives or counterarguments ($d = 0.22$, $p = 0.057$). Standard errors were clustered by group to account for within-group correlation, and the pattern also held under a linear regression. Full output is shown in Table~\ref{tab:app_usability_ols}.

Two other surveys asked about participants’ experience with the task. The first, a task-feedback survey, asked how well the discussion flowed, whether members pooled information and weighed the options, time management, mutual respect, and individual focus. The second, the NASA-TLX survey, measured cognitive load, including mental and temporal demand, performance, effort, and frustration. Responses to neither survey differed across conditions: no pairwise test on any item reached significance. Item-level results are shown in Figures~\ref{sifig:NASA} and~\ref{sifig:task_survey}.

\section*{Discussion}
Across 138 three-person groups completing a hidden profile hiring task, LLM facilitation increased the rate of correct consensus from 32.6\% to 56.5\%. Notably, the facilitator had access to neither the candidates’ attribute ratings nor the correct choice, which suggests that LLM facilitation can improve group accuracy even when privacy concerns or low trust in AI make sharing sensitive information undesirable. Our findings advance prior work showing that LLM facilitation can increase the amount of information shared within groups but does not reliably improve decision accuracy \cite{Alsobay2025-al}. This improvement may be due to the design of our facilitator, which targeted three well-established barriers to effective information pooling identified in the hidden profile literature: information overload, unequal participation, and premature convergence \cite{Lu2011_hidden_survey,Sohrab2015_HP_rev}. Each barrier was addressed with a separate intervention: summarizing the discussion, encouraging quieter members to contribute, and introducing a devil’s advocate argument when consensus formed early. Interventions were triggered by an LLM’s assessment of the need for each intervention type rather than on a fixed schedule. This design is consistent with recent literature identifying both the nature of interventions and the proactivity of LLMs as key determinants of successful LLM facilitation in group decision-making \cite{mao2024multiuserchatassistantmuca,Chiang2024-ly,butler2025nfw}.

Prior work on the hidden profile paradigm identifies two key components of decision-making: the sharing and the processing of information \cite{SchulzHardt2012_how_synergy}. Our transcript analyses suggest how improvements in both may have occurred. On the information-sharing side, groups in the individual-level condition repeated public attributes significantly less often than groups without facilitation, and groups in the group-level condition disclosed more private attributes about the correct candidate. These patterns map directly onto well-established findings from the hidden profile literature \cite{Stasser1985-rd,Lu2011_hidden_survey,Sohrab2015_HP_rev,MesmerMagnus2009_infosharing,SchulzHardt2012_how_synergy}, which makes them a possible explanation, though not causal evidence, for the improvement in decision accuracy. They are also in line with recent research finding that LLMs can increase information sharing in a similar group decision task \cite{Alsobay2025-al} as well as enhance topic diversity in a group brainstorming task \cite{cueing_the_crowd_dow,dow_2024}. On the information-processing side, both LLM-facilitated conditions produced more messages oriented toward reaching consensus. Participants in the group-level condition also produced more messages coordinating the decision process and discussed less nonspecific information about candidates. These findings might reflect a beneficial effect of LLM facilitation on groups’ collective attention, which previous research has identified as a key process underlying group decision-making \cite{Woolley2010-pk_evidence,Woolley2025-dz_opt_ci}. For example, prior work has shown that disruption of collective attention impairs the performance of virtual teams \cite{noshir_attention}. These results are also consistent with recent findings that AI can alter human–human communication and influence collective attention in human–AI teams \cite{riedl_spillover}.

Interestingly, groups in the individual-level condition showed a significant accuracy gain even though they did not significantly increase their disclosure of private attributes about the correct candidate. Exploratory evidence suggests that, although individual-level facilitation may not significantly improve this kind of information sharing at our sample size, it may be more effective at translating surfaced information into correct decisions (Table \ref{sitab:rational}). Once at least one participant has enough information to identify the correct candidate, individual-level facilitation might help groups convert that rational basis into a unanimous vote. In other words, individual-level LLM facilitation might help groups act correctly on more limited knowledge. One possibility is that individual-level facilitation changes how participants evaluate the information they receive. Previous research has shown that information running counter to participants’ initial preferences is discounted, and that information is more persuasive when corroborated by other participants~\cite{Greitemeyer2003_preference_consistent,Mojzisch2009_validation}. For example, the facilitator could reiterate information from others that runs against a participant’s initial preference, counteracting the tendency to discount it, while also validating information contributed by other members.

The usability surveys show a systematic preference for the group-level LLM condition. Participants rated the group-level facilitator higher on nearly every item, most strongly on the sense of collaborating with it. In open-ended responses, participants in the individual-level condition wished that ``the messages were universal info between all participants'' and disliked that the LLM ``had different responses for every group member,’’ whereas participants in the group-level condition described the facilitator as having ``helped organize ideas'' for the group, and one stated that ``it seemed to understand me more than my teammates did.’’ This result is consistent with previous qualitative findings that participants perceive a lack of transparency when messages are sent privately to individuals rather than to the group \cite{Do2022_groupvsprivate}. Notably, this preference did not carry over to how participants rated their group’s process or the cognitive load of the task, neither of which differed across conditions (including the no facilitation condition). These patterns offer some guidance for practical deployment: the two facilitation formats may suit different settings, and individual-level facilitation might be more effective for tasks in which a group would benefit from more personalized interventions.

There are several avenues for future work that could address the limitations stemming from our deliberately controlled setting. On the conceptual side, task difficulty is a known moderator of group synergy \cite{Almaatouq2021_difficulty}. In our task, members were not all biased toward the same wrong candidate. Future studies could align those initial preferences to make the task even harder, or move in the other direction and loosen the structure that makes hidden profiles so difficult. More broadly, the hidden profile task has a correct answer and splits information cleanly across members, which allowed us to measure how much information groups pooled. Many real decisions have neither feature, and whether a process-level LLM facilitator helps when there is no decisive fact to surface, or no correct answer to reach, remains an open question.

On the research design side, our threshold-based approach to the evaluator LLM could be improved with a more granular, empirically validated design for deciding when to intervene. Approaches from recent studies could be incorporated to improve the timing of interventions, for example by targeting specific types of conversational failure or by giving LLMs ``inner thoughts'' that model their ``motivation’’ to intervene \cite{proactive_proact,Yoo2026_proactive_readtheroom,Liu2025_proactive_inner}.

On the practical side, it remains to be seen whether the effects documented here hold outside the laboratory, in organizational decisions where group members know each other personally and have established group dynamics. Our groups comprised three strangers who met once over text, not the standing teams that make most real-life decisions, and it is unclear how the facilitator would perform with people who share complex social dynamics. In a setting involving professionals from the same company doing product idea generation, a previous experiment found that including an ``AI teammate'' improved individual performance to a level comparable to that of teams, but did not yield significant performance differences between teams with and without AI \cite{DellAcqua2026_teammate}. Finally, we did not compare our results with human facilitators. Previous research found no significant difference between LLM facilitation and facilitation by participants assigned to the facilitator role \cite{Alsobay2025-al}, but comparison with trained facilitators remains an open question.

Notwithstanding these limitations, our findings contribute to the growing literature on AI for group decision-making and teamwork \cite{Woolley2025-dz_opt_ci,Gonzalez2026_complementarity,Cui2024-dm-enhancedci,butler2025nfw,Burton2024-le-reshapeCi}. Our results show that LLM facilitation can improve group performance and may offer an efficient option when trained human facilitators are unavailable or prohibitively costly. Practically, they suggest that deploying LLMs as group facilitators, rather than as individual assistants, may be more acceptable to participants without sacrificing accuracy, at least in similar tasks. Critically, our LLM interventions were designed \emph{only} to support the human decision-making process, without knowledge of the correct answer and with the explicit goal of preserving full human agency. This contrasts with approaches that rely on LLM output more directly, such as having groups approve consensus statements drafted by LLM systems \cite{Tessler2024_habermas}. Even so, adopting this kind of technology carries risks worth stressing: groups could grow accustomed to process-preserving interventions, potentially eroding their ability to organize themselves without such tools. Our work promotes a more thoughtful use of AI assistants in real group discussions, one that grounds interventions in relevant theory and helps groups decide better without deciding for them.

\section*{Methods}\label{sec_methods}

\paragraph{Experimental procedure}\label{subsec_participants}
We recruited participants using Prolific, screening for an approval rate above 97.5\% \cite{prolific}. We excluded 42 groups in which at least one participant did not cast a final vote, 4 groups in which one participant sent no messages, and 2 groups with fewer than 10 messages in total. We dropped these groups in accordance with our preregistered exclusion rules (see preregistration \#280,167 on AsPredicted). This left complete data for 138 groups comprising 414 participants. We collected demographics in a pre-task survey, including age, gender, education, race, and familiarity with AI (see Table \ref{sitab:participant-demographics} for the demographic composition of our sample). Participants received a base payment of \$2.80 and a potential bonus of \$1.50 each if their group selected the correct candidate. The experiment was approved by the Northwestern University IRB (STU00223965) and the Purdue University IRB (2025-771).

All participants gave informed consent. They then read the task instructions and completed two comprehension checks: a true/false question confirming they understood that group members could hold different information about the candidates, and an example confirming they understood that the correct candidate was the one with the highest number of exceptional attributes. Participants could not continue until they answered both checks correctly. Each participant then received their private information sheet and submitted an initial choice before entering the three-person text chat phase. The information sheet remained available throughout the discussion. Groups discussed the candidates for at least five minutes. In the two LLM conditions, the facilitator monitored the chat and could intervene during the discussion according to the condition assignment (Figure \ref{fig:llm_logic}). After the discussion, each participant independently submitted a final candidate choice. Participants then completed two or three post-task surveys, depending on the condition.

\typeout{}
\bibliography{sn-bibliography}

\backmatter
\bmhead{Acknowledgments}
We thank the members of the LINK Lab at Northwestern University (Katherine O’Toole, Maalvika Bhat, Haohan Shi, and Nicholas Ornstein) for their feedback and suggestions during the implementation and writing of this study. This work was partially supported by NSF CAREER Award IIS-1943506 and NSF grants IIS-2229876 and IIS-2340209.
\clearpage

\section*{Supplementary Information}

\setcounter{section}{0}
\renewcommand{\thesection}{S\arabic{section}}
\renewcommand{\thefigure}{S\arabic{figure}}
\renewcommand{\thetable}{S\arabic{table}}
\renewcommand{\theequation}{S\arabic{equation}}

\setcounter{figure}{0}
\setcounter{table}{0}
\setcounter{equation}{0}

\newcommand{\sitext}[2]{%
  \refstepcounter{section}%
  \section*{Text \thesection: #1}\label{#2}%
}

\begin{table}[h]
\centering
\caption{Three candidates (A, B, C) are rated on ten attributes (average/good/exceptional). Four attributes are public (i.e, every participant has access to them); the remaining six are private, each pair held by a single participant. Across the full set of information, candidate A is the correct choice, with four exceptional attributes against three each for candidates B and C. Only one of A's four exceptional attributes is among the attributes available to all participants; and the other three are divided across members' private information. Private attributes are arranged so that, before discussion, each participant's own information favors a different candidate: one favors A, one B, and one C. Identifying A as the correct candidate requires participants to disclose and combine the information they hold privately, and to reach a consensus on the group choice.}
\label{sitab:attributes}
\renewcommand{\arraystretch}{1.2}
\begin{tabular}{llccc}
\hline
\textbf{Attribute} & \textbf{Held by} & \textbf{Candidate A} & \textbf{Candidate B} & \textbf{Candidate C} \\
\hline
Publications        & All members (public) & Good                 & \textbf{Exceptional} & Average              \\
Citations           & All members (public) & \textbf{Exceptional} & Good                 & Good                 \\
Editorial / Service & All members (public) & Average              & Average              & \textbf{Exceptional} \\
Conference          & All members (public) & Good                 & Good                 & \textbf{Exceptional} \\
\hline
Mentorship          & Participant 1        & \textbf{Exceptional} & Average              & Average              \\
Teaching            & Participant 1        & \textbf{Exceptional} & Good                 & Good                 \\
\hline
Funding             & Participant 2        & Good                 & \textbf{Exceptional} & Average              \\
Interdisciplinarity & Participant 2        & Good                 & \textbf{Exceptional} & Good                 \\
\hline
Collaborations      & Participant 3        & \textbf{Exceptional} & Average              & Good                 \\
Research Coverage   & Participant 3        & Good                 & Average              & \textbf{Exceptional} \\
\hline
\end{tabular}
\end{table}

\begin{figure}[H]
    \centering
    \includegraphics[width=1\textwidth]{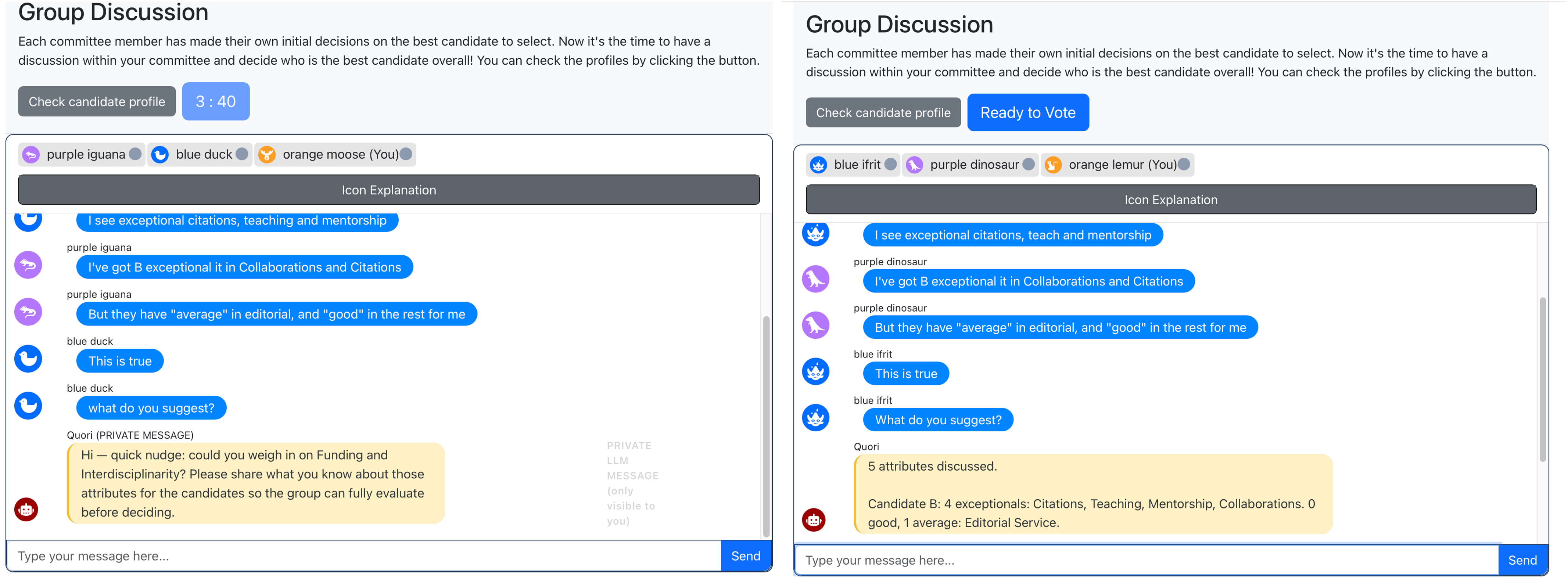}
    \caption{Chat example for individual (left) and group (right) LLM facilitation.}
    \label{sifig:chatbox}
\end{figure}

\sitext{Prompts for the evaluator LLM}{si:prompts_evaluator} 
\begin{verbatim}
You are **Intervention Analyzer**.  
Mission: Inspect the latest slice of a 3 person group‑discussion 
transcript and predict how *helpful* three facilitator interventions would 
be **right now**. 
The transcript includes messages from an AI assitant (quori)

**Input**  
`<TRANSCRIPT>` - raw lines exactly as posted (newest last).  

**Interventions to score**  
1. **summarization** – concise synthesis of key points.  
2. **nudging_participants** – prompt quiet/off-topic voices.  
3. **devils_advocate** – challenge consensus or assumptions.

**Output you will output a json nothing else**  
{
  "summarization": {"score": 0, "justification": ""},
  "nudging": {"score": 0, "justification": ""},
  "devils_advocate": {"score": 0, "justification": ""}
}


`score' = int 0 - 100.  
`justification' less than 50 words.

Heuristics for assigning scores
Choose a numeric value on these bands:

Strong need: 80-100
Moderate need: 60-70
Neutral: 40-50
Counter-productive: 10-30
Use the cues below .

1. Summarization

Raise when

Several distinct items or sub-topics are being discussed in parallel.
Speakers repeat points or ask "what have we dicussed?"

Lower when

A summarization was provided recently, for example during the 
last 5 messages.
Discussion is focused on one well-understood issue.
There has been a been an intervention by an assistant summarizing 
the discussion in recent messages. 
Discussion has only started

2. Nudging Participants

Raise when

One or two voices dominate; at least one participant hasn’t spoken
in the last 8-10 lines.
Conversation drifts off the stated agenda or stalls 
(long pauses, short "yeah" replies).
A participant has only contributed overall information but has 
not explicitly mentioned specific attribute names. 
Remember there are 3 participants in the group.

Lower when

All participants have contributed substantive messages in
recent turns.
There has been a nudging of a the participant in recent messages.
Discussion has only started (less than 5 messages).
There was nudge recently, for example during the last 5 messages.

3. Devil’s Advocate

Raise when

Rapid agreement or unanimous approval between the 3 participants 
appears without exploring alternatives.
No objections or critical questions are present.

Lower when

There was a Devil’s advocate recently, for example during the 
last 5 messages.
Active disagreement or diverse viewpoints already exist.
Topic is still exploratory (no concrete proposal to challenge).
There has been a devils advocate intervention by an assistant 
in recent messages.
Discussion has only started.
\end{verbatim}

\sitext{Prompts for the generator LLM}{si:prompts_generator} 

\begin{Verbatim}[
  breaklines=true,
  breakanywhere=true,
  fontsize=\small
]
You are an assistant named quori to a group engaged in a collaborative decision-making task between 3 persons. The group is evaluating candidates for a professor position based on partial information distributed among them. Guide the discussion to ensure thorough examination using both shared and private evaluation criteria. You should:

Track the conversation.
Prompt deeper discussion as needed.
Ensure all members consider both shared and private criteria.

Evaluation Criteria:

Shared: {public_information}.
Private:
 {role_id_1}: Mentorship, Teaching
 {role_id_2}: Funding, Interdisciplinarity
 {role_id_3}: Collaborations, Research Coverage

Maintain a conversational, direct, and open-ended style to foster dialogue.
Ensure prompts are concise, relevant to criteria, and integrate all information as discussions progress.
Procedure:

You will be prompted to perform one of three following tasks, you should perform only the correct task:

1 Summarization:

Review the transcript line-by-line to identify attributes discussed for each candidate and their values.
Do not mention what information is private or public at all.

Guide of Output:

First: Start with providing the number of specific attributes that have been discussed, i.e. how many attributes of the evaluation criteria have been explicitly mentioned. Provide a concise summary of attributes discussed.

Example of output: 
Candidate A: 2 exceptionals: attribute 1, attribute 2. 1 good: attribute 3, 0 average.
Candidate B: 1 exceptionals: attribute 3. 1 good: attribute 6, 0 average.
Candidate C: 2 exceptionals: attribute 4, attribute 7. 2 good: attribute 1, attribute 2, 1 average: attribute 3.


2 Nudging Participant:

Select a participant to nudge. Specifically check the messages from the 3 participants and select the one that has not contributed meaningfully in the last messages. For example: if a participant has not contributed information regarding specific attributes. 

Provide a brief, friendly nudge to this participant to discuss attributes or information they might have overlooked or not fully explored yet. 

Do not nudge participants that have contributed recently and meaningfully.
Do not mention what information is private or public at all.
Do not sumarize the attributes that have been discussed.

Guide of Output:

Include the teammate's name, and a polite nudge to increase the participation of the participant (either suggesting new information or providing input to the previous messages). Make sure the response is not longer than 75 words. 

3 Devils Advocate:

Identify the current consensus of the best applicant and act as a devil's advocate by gently but clearly challenging the group's current consensus. Introduce thoughtful questions or highlight overlooked strengths of less favored candidates encouraging participants to reconsider or more deeply analyze alternative perspectives before finalizing their decision.
Focus solely on the factual differences
Do not mention what information is private or public at all.
Do not summarize attributes that have been discussed. 

Guide of Output:

State what the current consensus is, select another candidate that can rival the consensus, provide strong reasons for the rival candidate. Make sure the response is not longer than 75 words. 

Constraints for the 3 tasks:

Do not fabricate attribute values. If there are no attributes discussed yet, do not include any.
Ensure clarity and precision in summarizing discussions and suggesting next steps.
Do not include judgment on what attributes might be important. Only list them
Make sure you don't make claims regarding which attributes can be more important. 
Do not include what type of intervention it is. 
Make sure the message is natural and does not appear to artificial. 
Make sure that you use the name of the attributes and not attribute numbersor letters eg. attribute A or attribute 1. 
Do not mention what information is private or public at all.
Make all replies less than 75 words.


\end{Verbatim}

\begin{Verbatim}[
  breaklines=true,
  breakanywhere=true,
  fontsize=\small
]
You are an assistant named quori to {participant_name} who is engaged in a collaborative decision-making task between 3 persons. The group is evaluating candidates for a professor position based on partial information distributed among them. Guide the discussion to ensure thorough examination using both shared and private evaluation criteria. You should:

Track the conversation.
Prompt deeper discussion as needed.
Ensure all members consider both shared and private criteria.

Evaluation Criteria:

Shared: {shared_information}.
Private:
 {participant_name}: {private_information}


Maintain a conversational, direct, and open-ended style to foster dialogue.
Ensure prompts are concise, relevant to criteria, and integrate all information as discussions progress.
Procedure:

You will be prompted to perform one of three following tasks, you should perform only the correct task:

1 Summarization:

Review the transcript line-by-line to identify attributes discussed for each candidate and their values.
Compare this with the available evaluation criteria. 
Do not mention what information is private or public at all.


Guide of Output:

First: Start with providing the number of specific attributes that have been discussed i.e. how many attributes of the evaluation criteria have been explicitly mentioned. Provide a concise summary of attributes discussed.

Address {participant_name} as 'you' (second-person); do not mention their name.
Do not infer or invent any information.

Example of output: 
Candidate A: 2 exceptionals: attribute 1, attribute 2. 1 good: attribute 3, 0 average.
Candidate B: 1 exceptionals: attribute 3. 1 good: attribute 6, 0 average.
Candidate C: 2 exceptionals: attribute 4, attribute 7. 2 good: attribute 1, attribute 2, 1 average: attribute 3.


2 Nudging Participant:

Provide a brief, friendly nudge to encourage participant {participant_name} to discuss attributes or information they might have overlooked or not fully explored yet. 

Do not mention what information is private or public at all.
Do not sumarize the attributes that have been discussed.

Guide of Output:


Provide a brief, friendly nudge to this participant to increase their participation. Address {participant_name} as 'you' (second-person); do not mention their name. Make sure the response is not longer than 75 words. 



3 Devils Advocate:

Identify the current consensus of the best applicant and act as a devil's advocate by gently but clearly challenging the group's current consensus. Introduce thoughtful questions or highlight overlooked strengths of less favored candidates encouraging participants to reconsider or more deeply analyze alternative perspectives before finalizing their decision. 
Focus solely on the factual differences
Do not mention what information is private or public at all.
Do not sumarize the attributes that have been discussed.


Guide of Output:

State what the current consensus is, select another candidate that can rival the consensus, provide strong reasons for the rival candidate by focusing on {participant_name}'s attributes . Address {participant_name} as 'you' (second-person); do not mention their name. Make sure the response is not longer than 75 words. 




Constraints for the 3 tasks:

Do not fabricate attribute values. If there are no attributes discussed yet, do not include any.
Ensure clarity and precision in summarizing discussions and suggesting next steps.
Do not include judgment on what attributes might be important. Only list them
Make sure you don't make claims regarding which attributes can be more important. 
Do not include what type of intervention it is. 
Make sure the message is natural and does not appear to artificial. 
Make sure that you use the name of the attributes and not attribute numbers. 
Do not mention what information is private or public at all.
Make all replies less than 75 words.
\end{Verbatim}

\sitext{Codebook for detecting communicative functions}{si:codebook}

Our approach builds on Bales' utterance-level functional analysis, which focuses on the communicative function of an utterance beyond its informational content \cite{bales1950interaction}. This includes categories such as agreement or stating an opinion. Bales' framework was later extended to group decision-making \cite{gouran1996functional} and team coordination behavior \cite{kolbe2011microco}. Similar approaches have since been applied to hidden profile tasks to identify agreement with a potential group choice, information elicitation, and process coordination \cite{thurmer2018hidden}. In our coding scheme, we adjusted these categories to match our specific faculty hiring task and the context of a brief three-person online chat. This led us to five variables in our final codebook.

\begin{itemize}
\item \textbf{Consensus}: Messages showing agreement with a group choice (e.g., ``Yes, A seems like a good choice'', ``I think the case for B is very strong'' ).

\item \textbf{Coordination}: Messages managing how the group would make its decision (e.g., ``Who do we settle on?'', ``How do we vote if that's the case?'', ``With all the categories compiled, which has the highest amount of exceptionals?'').

\item \textbf{Elicitation}: Messages eliciting information from teammates (e.g., ``What was A in research?'', ``How many exceptionals did you get for each candidate?'', ``Do you recall in which categories?'').

\item \textbf{Contribution of specific information}: Messages stating a concrete attribute of a candidate (e.g., ``A is exceptional in mentorship and teaching'', ``C was exceptional in research, conference, and editorial'', ``A has exceptional collaborations and citations'').

\item \textbf{Contribution of nonspecific information}: Messages contributing general information without naming a concrete attribute (e.g., ``I chose candidate A, exceptional in 3 areas'', ``Candidate C has the most exceptionals'').

\end{itemize}

\sitext{Usability survey items}{si:sec_survey_questions}

Participants in the two LLM conditions (individual- and group-level LLM facilitation) rated the following eight items on a five-point Likert scale (1: Strongly Disagree to 5: Strongly Agree) after the task. The first three items were created by us to assess the effectiveness of our specific interventions. The next three items were adapted from previous research to assess participants' interactions with the LLM facilitator \cite{Chiang2024-ly,Yuan2022_survey}. The final two items were included to evaluate participants' behavioral intentions and were adapted from the Word of Mouth Behavioral scale \cite{Zeithaml1996_reccomend_scale} and the Unified Theory of Acceptance and Use of Technology \cite{Venkatesh2003_survey_use}:

\begin{enumerate}
    \item The LLM facilitator did a good job summarizing information when needed to reach a decision.
    \item The LLM facilitator did a good job encouraging participants to share information to reach a decision.
    \item The LLM facilitator raised alternative perspectives or provided useful counterarguments to reach a decision.
    \item I feel like I was collaborating with the LLM facilitator during the task.
    \item I'm satisfied with the assistance provided by the LLM facilitator in completing the tasks.
    \item I'm pleased with the quality of the LLM facilitator in completing the tasks.
    \item I would recommend the LLM facilitator to my friends if they need to complete similar tasks.
    \item If given the option, I would use the LLM facilitator to assist me with completing similar decision-making tasks in the future.
\end{enumerate}

Participants could also leave optional free-text comments about their experience with the facilitator. We highlighted some of these comments in the \textit{Discussion}.

\begin{figure}[H]
    \centering
    \includegraphics[width=1\textwidth]{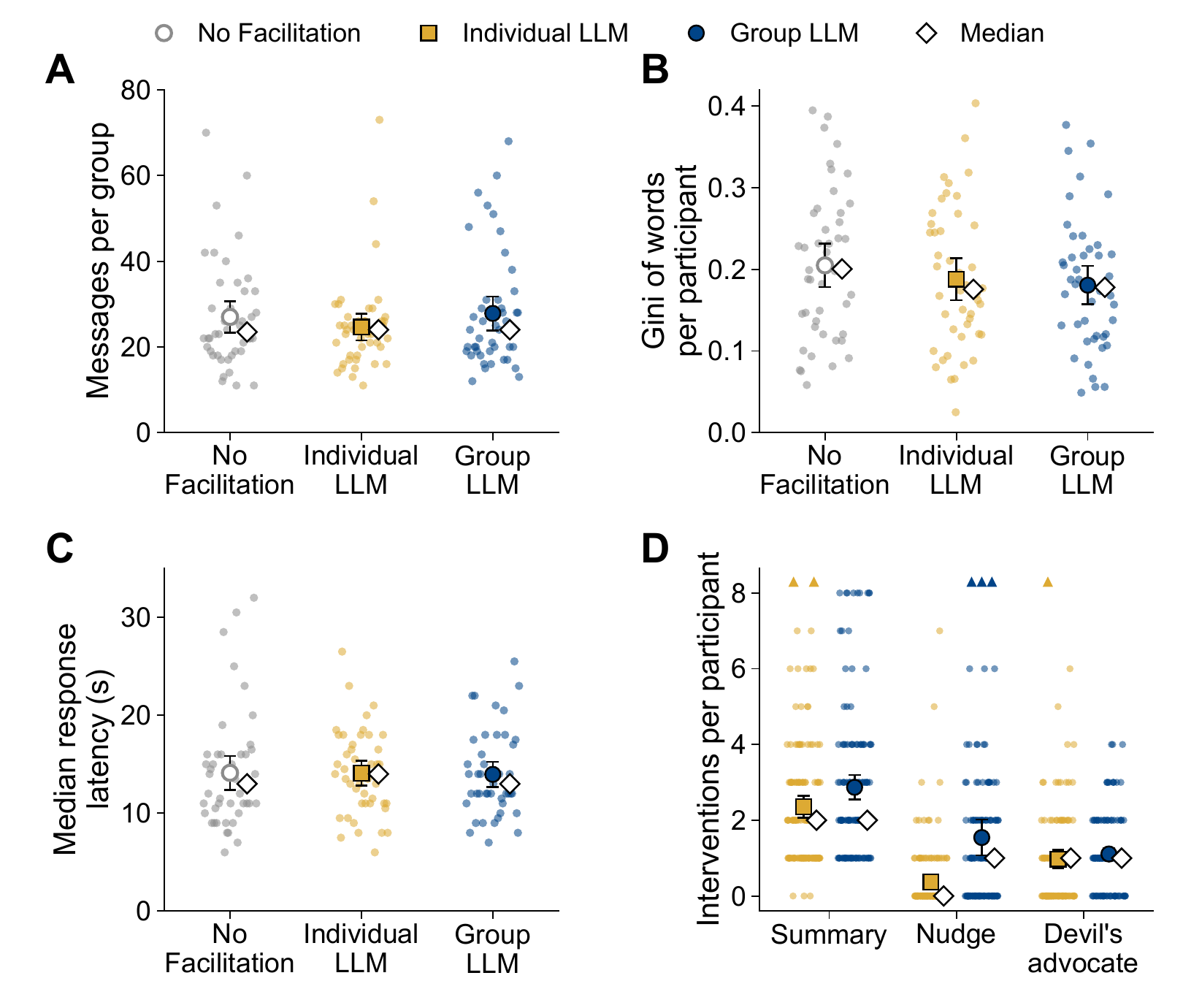}
    \caption{Conversation dynamics and intervention delivery by condition. Each point is one group; dots show the mean ($\pm$95\% CI) and diamonds indicate the median. (A)~Number of messages per group, (B)~Gini coefficient of words per participant, and (C)~median response latency between messages did not differ across conditions, indicating that the experimental conditions were comparable in how much, how evenly, and how quickly groups conversed. Groups exchanged an average of 26.5 messages over a median duration of 7.7 minutes, with messages containing an average of 8 words. The distribution of participation within groups was also similar across conditions, with a mean Gini coefficient of $\approx$0.19 indicating relatively equal participation among participants. The median response latency between messages was 14 seconds.
    (D)~Number of interventions of each type that participants were exposed to. Almost all participants received at least one intervention (99\% in the individual-level LLM facilitation condition and 100\% in the group condition), so the manipulation reached nearly everyone. The three types of interventions, however, reached different shares of participants. Summarization reached nearly all participants in both conditions (98\% in the individual condition, 100\% in the group condition). The percentage of participants exposed to the devil's advocate intervention rose from 56\% in the individual condition to 76\% in the group condition, which is unsurprising given that devil's advocate is naturally conceptualized at the group level \cite{Chiang2024-ly}. Nudging, by contrast, has been defined at the individual level; the apparent increase in its frequency (from 22\% in the individual
    condition to 59\% in the group condition) is explained by the fact that, by design, the same nudge was seen by all three participants in a group.
    Together, these checks show the conditions were well matched on how groups talked, and that facilitation reached nearly every participant, with exposure to specific intervention types varying across LLM facilitation conditions.}
    \label{sifig:con_characteristics}
\end{figure}

\begin{table}[htbp]
\centering
\small
\setlength{\tabcolsep}{4pt}
\caption{Logistic regression of group success on facilitation condition}
\label{tab:app_accuracy_logit}
\begin{tabular}{lcc}
\toprule
 & (1) Condition & (2) Any LLM \\
\midrule
Individual LLM & 2.69 &  \\
 & [1.15, 6.28] &  \\
 & \textit{p} = 0.022 &  \\[3pt]
Group LLM & 2.69 &  \\
 & [1.15, 6.28] &  \\
 & \textit{p} = 0.022 &  \\[3pt]
Any LLM (pooled) &  & 2.69 \\
 &  & [1.28, 5.64] \\
 &  & \textit{p} = 0.009 \\[3pt]
Constant (baseline odds) & 0.48 & 0.48 \\
 & [0.26, 0.90] & [0.26, 0.90] \\
 & \textit{p} = 0.021 & \textit{p} = 0.021 \\[3pt]
\midrule
Observations (groups) & 138 & 138 \\
Log-likelihood & $-$92.03 & $-$92.03 \\
Pseudo $R^2$ (McFadden) & 0.037 & 0.037 \\
LR $\chi^2$ (df) & 7.14 (2) & 7.14 (1) \\
LR test \textit{p} & 0.028 & 0.008 \\
\bottomrule
\end{tabular}

\smallskip
\begin{minipage}{0.95\linewidth}\footnotesize
\textit{Note.} Entries are odds ratios with 95\% confidence intervals in brackets and two-sided Wald \textit{p}-values. No Facilitation is the reference category. Logistic regression, one observation per group. Outcome: the group's formal vote was unanimous for the best candidate. Model (2) pools both LLM arms.
\end{minipage}
\end{table}

\begin{figure}[H]
    \centering
    \begin{minipage}[t]{0.49\textwidth}
        \centering
        \includegraphics[width=\linewidth]{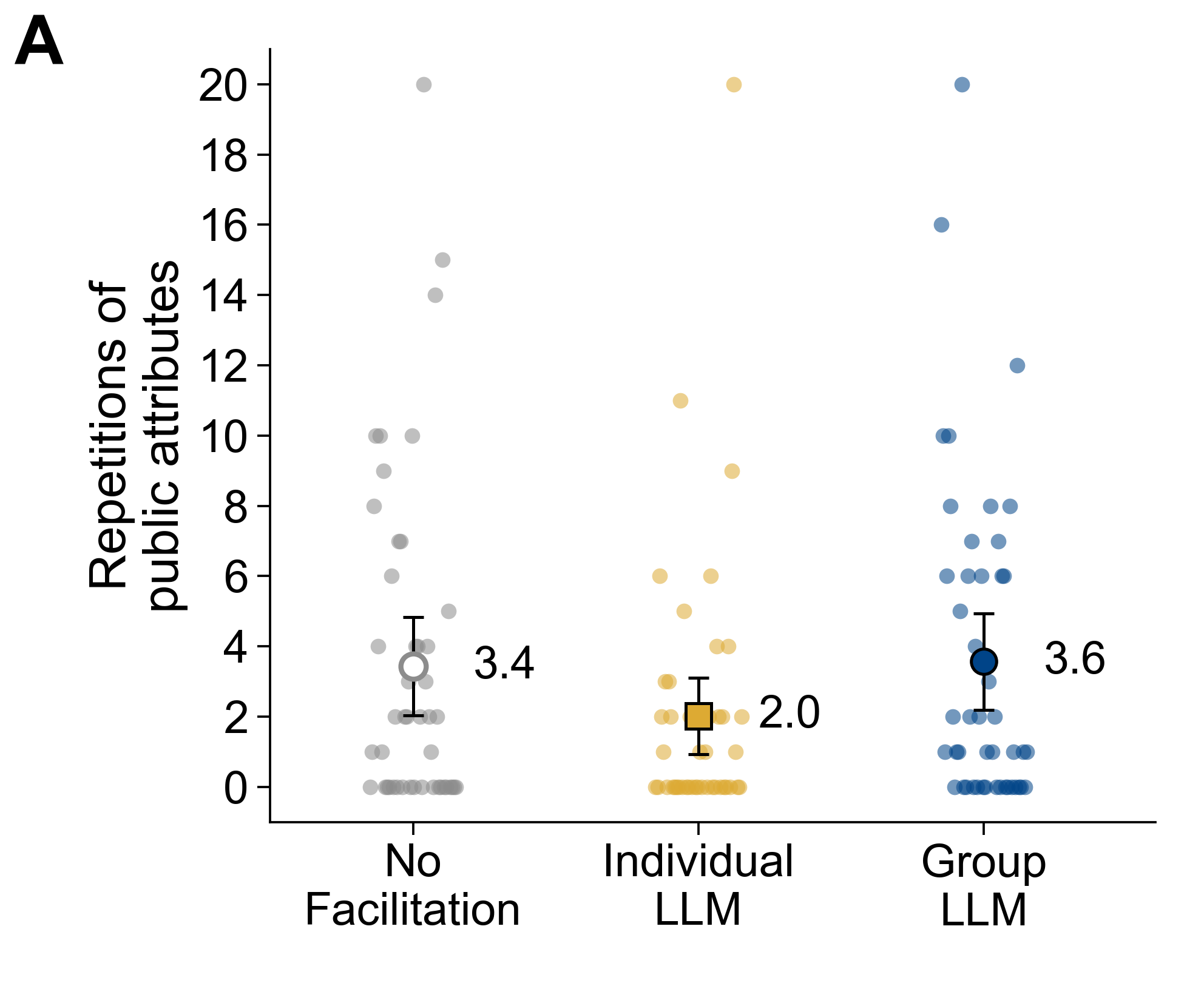}
    \end{minipage}
    \hfill
    \begin{minipage}[t]{0.49\textwidth}
        \centering
        \includegraphics[width=\linewidth]{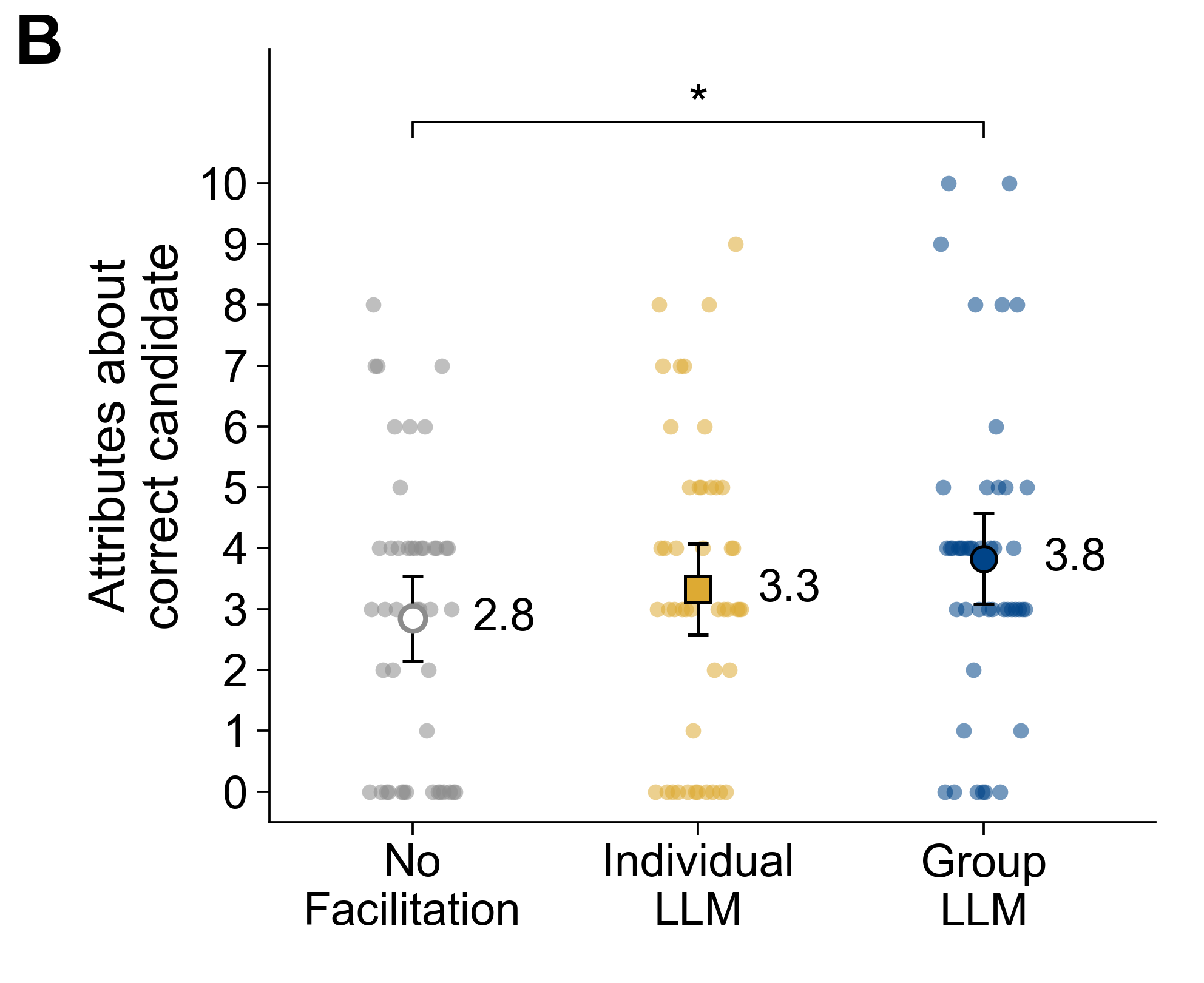}
    \end{minipage}
    \caption{Information sharing by condition. \textbf{(A)} Repetition of public attributes during group discussion: repetitions per group by condition (mean with 95\% CI). \textbf{(B)} Total number of attributes about the correct candidate, both private and public, mentioned during group discussion, by condition.}
    \label{sifig:information_disclousre_and_repetition_appendix}
\end{figure}

\begin{table}[htbp]
\centering
\small
\setlength{\tabcolsep}{4pt}
\caption{Negative binomial regression of public-fact repetition}
\label{tab:app_repetition_nb}
\begin{tabular}{lc}
\toprule
 & Public-fact repeats \\
\midrule
Individual LLM & 0.48 \\
 & [0.29, 0.79] \\
 & \textit{p} = 0.004 \\[3pt]
Group LLM & 0.62 \\
 & [0.39, 1.01] \\
 & \textit{p} = 0.056 \\[3pt]
Distinct facts surfaced & 1.21 \\
 & [1.15, 1.28] \\
 & \textit{p} $<$ 0.001 \\[3pt]
Human messages & 1.03 \\
 & [1.01, 1.05] \\
 & \textit{p} $<$ 0.001 \\[3pt]
Constant (baseline rate) & 0.21 \\
 & [0.10, 0.41] \\
 & \textit{p} $<$ 0.001 \\[3pt]
\midrule
Observations (groups) & 138 \\
Log-likelihood & $-$253.62 \\
Pseudo $R^2$ (McFadden) & 0.150 \\
LR $\chi^2$ (df) & 89.62 (4) \\
LR test \textit{p} & $<$ 0.001 \\
Dispersion $\alpha$ [95\% CI] & 0.78 [0.44, 1.13] \\
\bottomrule
\end{tabular}

\smallskip
\begin{minipage}{0.95\linewidth}\footnotesize
\textit{Note.} Entries are incidence rate ratios with 95\% confidence intervals in brackets and two-sided Wald \textit{p}-values. No Facilitation is the reference category. Outcome: mentions of public facts beyond the first mention of each fact (human messages only).
\end{minipage}
\end{table}

\begin{table}[htbp]
\centering
\footnotesize
\setlength{\tabcolsep}{4pt}
\caption{Negative binomial regressions of communicative function message counts on condition}
\label{tab:app_speech_acts_nb}
\begin{tabular}{lccccc}
\toprule
 & \shortstack{Elicit\\information} & \shortstack{Non-specific\\information} & \shortstack{Specific\\information} & \shortstack{Consensus\\messages} & \shortstack{Coordination\\messages} \\
\midrule
Individual LLM & 0.87 & 0.87 & 0.90 & 1.57 & 1.19 \\
 & [0.64, 1.19] & [0.64, 1.18] & [0.66, 1.21] & [1.16, 2.11] & [0.82, 1.71] \\
 & \textit{p} = 0.388 & \textit{p} = 0.367 & \textit{p} = 0.478 & \textit{p} = 0.003 & \textit{p} = 0.361 \\[3pt]
Group LLM & 0.82 & 0.65 & 1.07 & 1.59 & 1.54 \\
 & [0.61, 1.12] & [0.48, 0.89] & [0.80, 1.44] & [1.18, 2.12] & [1.10, 2.15] \\
 & \textit{p} = 0.215 & \textit{p} = 0.007 & \textit{p} = 0.651 & \textit{p} = 0.002 & \textit{p} = 0.012 \\[3pt]
Group LLM vs Individual LLM & 0.94 & 0.75 & 1.19 & 1.01 & 1.30 \\
 & [0.69, 1.30] & [0.54, 1.03] & [0.88, 1.61] & [0.78, 1.31] & [0.93, 1.81] \\
 & \textit{p} = 0.725 & \textit{p} = 0.076 & \textit{p} = 0.246 & \textit{p} = 0.940 & \textit{p} = 0.123 \\[3pt]
Human Messages & 2.81 & 2.32 & 2.70 & 2.00 & 2.76 \\
 & [2.07, 3.81] & [1.68, 3.20] & [2.01, 3.63] & [1.53, 2.62] & [2.00, 3.80] \\
 & \textit{p} $<$ 0.001 & \textit{p} $<$ 0.001 & \textit{p} $<$ 0.001 & \textit{p} $<$ 0.001 & \textit{p} $<$ 0.001 \\[3pt]
Constant & 0.09 & 0.32 & 0.17 & 0.17 & 0.05 \\
 & [0.03, 0.25] & [0.11, 0.93] & [0.06, 0.47] & [0.07, 0.42] & [0.01, 0.14] \\
 & \textit{p} $<$ 0.001 & \textit{p} = 0.036 & \textit{p} $<$ 0.001 & \textit{p} $<$ 0.001 & \textit{p} $<$ 0.001 \\[3pt]
\midrule
Observations (groups) & 138 & 138 & 138 & 138 & 138 \\
Log-likelihood & $-$250.24 & $-$326.47 & $-$332.73 & $-$231.78 & $-$207.42 \\
Pseudo $R^2$ (McFadden) & 0.073 & 0.043 & 0.057 & 0.072 & 0.087 \\
LR $\chi^2$ (df) & 39.63 (3) & 29.21 (3) & 40.18 (3) & 35.78 (3) & 39.77 (3) \\
LR test \textit{p} & $<$ 0.001 & $<$ 0.001 & $<$ 0.001 & $<$ 0.001 & $<$ 0.001 \\

\bottomrule
\end{tabular}

\smallskip
\begin{minipage}{0.95\linewidth}\footnotesize
\textit{Note.} Entries are incidence rate ratios with 95\% confidence intervals in brackets and two-sided Wald \textit{p}-values. No Facilitation is the reference category. Negative binomial regressions, one observation per group; each outcome is the number of human messages with the specific communicative function. 
\end{minipage}
\end{table}

\FloatBarrier
\begin{table}[htbp]
\centering
\small
\setlength{\tabcolsep}{4pt}
\caption{OLS regressions of facilitator-usability ratings on condition}
\label{tab:app_usability_ols}
\begin{tabular}{lccc}
\toprule
Item & Group LLM & 95\% CI & \textit{p} \\
\midrule
Summaries & 0.25 & [0.01, 0.50] & 0.044 \\
Encouraged sharing & 0.43 & [0.15, 0.72] & 0.003 \\
Raised alternatives & 0.25 & [$-$0.01, 0.51] & 0.057 \\
Felt collaborative & 0.56 & [0.25, 0.86] & $<$0.001 \\
Satisfaction & 0.37 & [0.10, 0.64] & 0.007 \\
Quality & 0.38 & [0.12, 0.63] & 0.004 \\
Would recommend & 0.36 & [0.11, 0.62] & 0.005 \\
Use again & 0.31 & [0.05, 0.57] & 0.017 \\
\bottomrule
\end{tabular}

\smallskip
\begin{minipage}{0.95\linewidth}\footnotesize
\textit{Note.} Each row is a separate OLS regression of the item rating (1--5 Likert) on condition, one observation per participant ($N = 276$ participants in 92 groups). Entries are the Group LLM coefficient, i.e.\ the difference in mean rating relative to Individual LLM (the reference). Confidence intervals and \textit{p}-values use cluster-robust standard errors by group.
\end{minipage}
\end{table}

\FloatBarrier

\begin{figure}[H]
    \centering
    \includegraphics[width=1\textwidth]{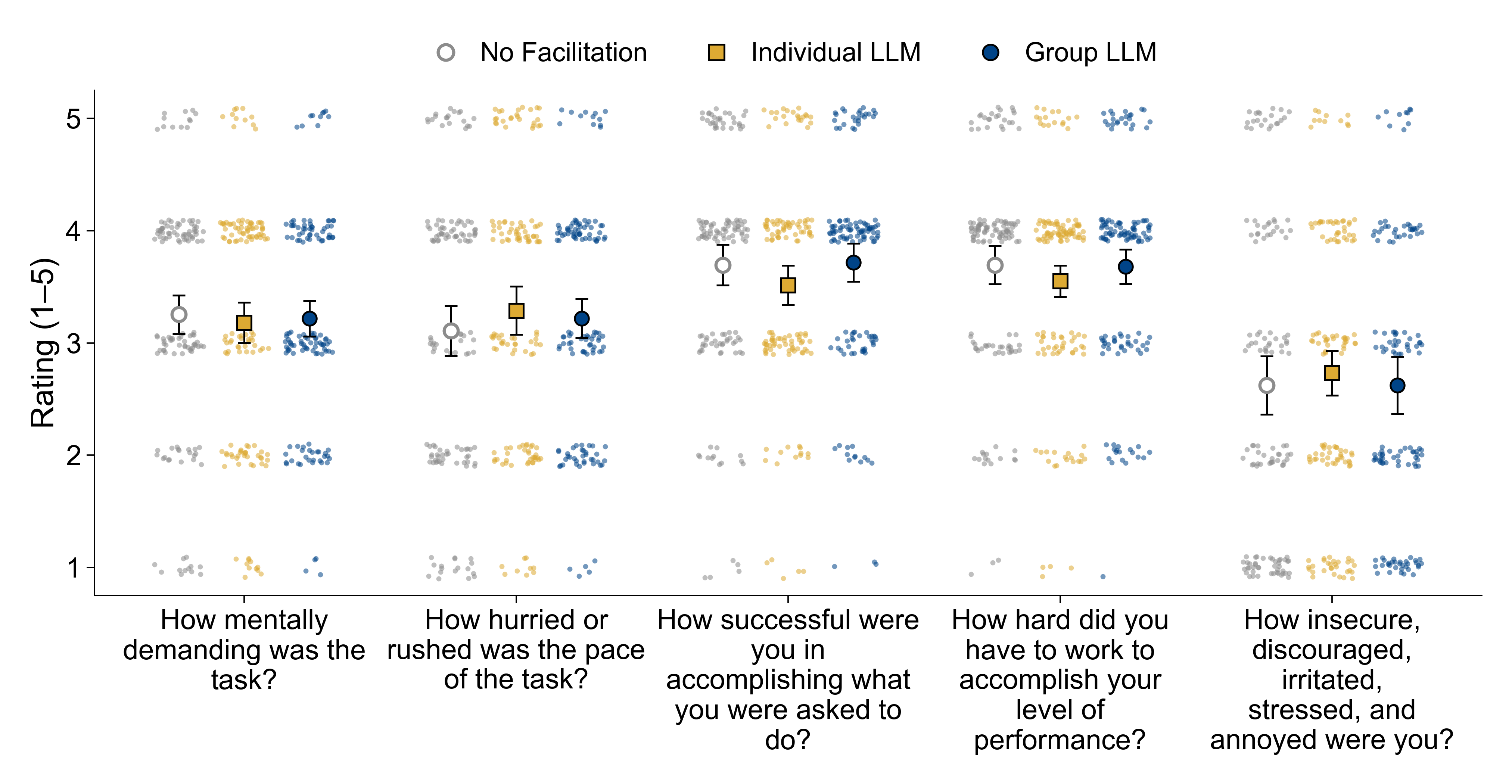}
    \caption{NASA-TLX ratings measuring perceived task workload, by condition. Dots are condition means with 95\% confidence intervals; faded dots are individual participant responses (jittered). No item differed significantly across conditions.}
    \label{sifig:NASA}
\end{figure}

\begin{figure}[H]
    \centering
    \includegraphics[width=1\textwidth]{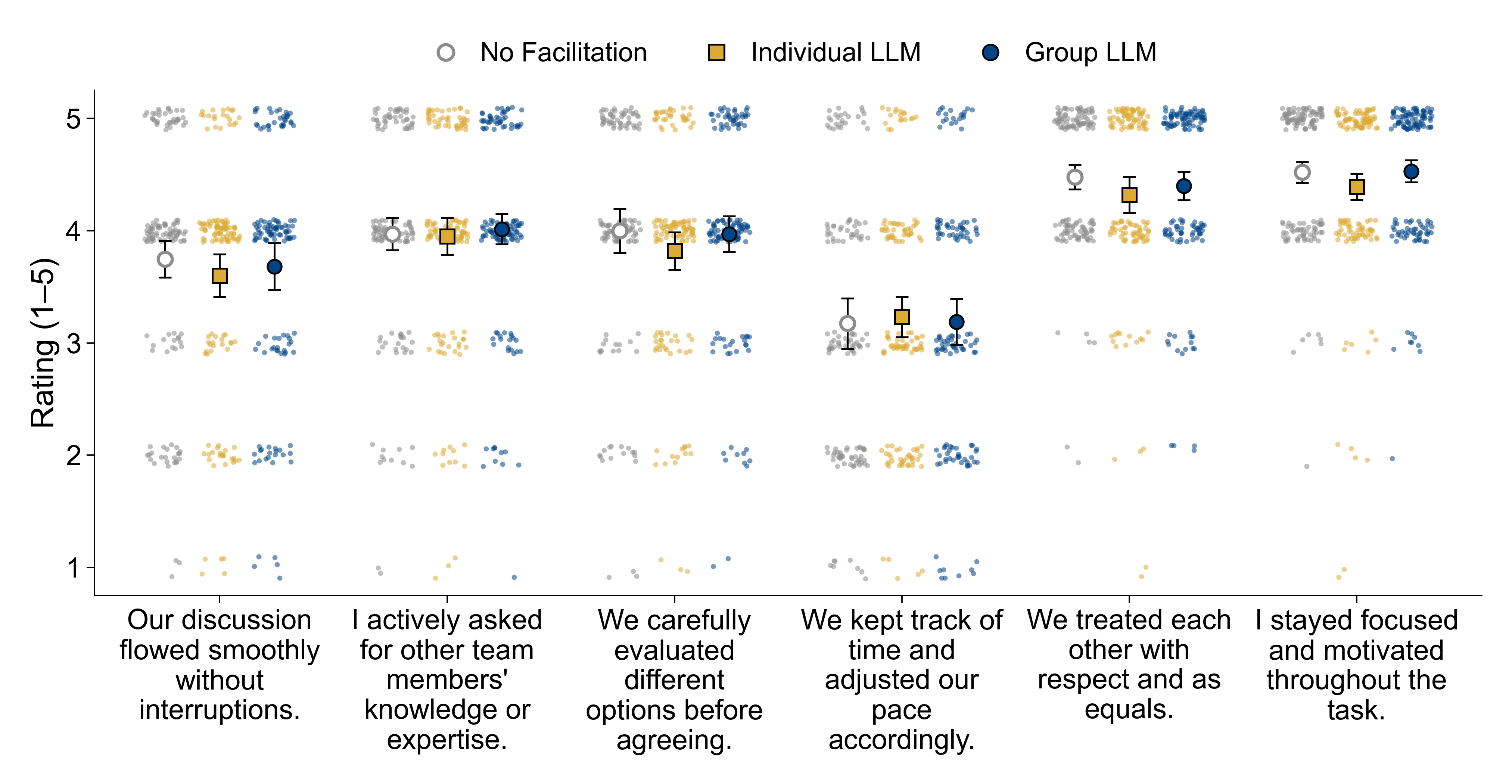}
    \caption{Task-feedback ratings of the group's process by condition. Dots are condition means with 95\% confidence intervals; faded dots are individual participant responses (jittered). No item differed significantly across conditions.}
    \label{sifig:task_survey}
\end{figure}

\begin{table}[h]
\caption{Percentage of groups that voted unanimously for the correct candidate A, by the number of participants with A as their rational choice, 0  (rows) and by experimental condition (columns). For each participant, A is their rational choice when---given their own information and everything surfaced during the discussion---A has strictly more visible exceptional attributes than either of the other two candidates. A participant can see an exceptional attribute if it was in the information they received pre-discussion, or if it was surfaced by the group. For each condition, we grouped all 46 discussions by the number of participants for whom A was the rational choice after discussion. Each cell reports two numbers in parentheses: the number of groups with that standing after discussion (second number), and the number of those groups that actually voted for A (first number). The percentage shown in each cell is the number of groups that voted for A divided by the number of groups with A as their rational choice, i.e, the decision accuracy given that standing. In our sample, the group-level LLM facilitation condition was best at surfacing information about the correct candidate: in half of the groups, two or three participants had access to all of A's exceptional attributes. The individual-level LLM facilitation condition, by contrast, was better at converting groups' more limited knowledge of A's exceptional attributes into the correct decision in nearly all groups where the correct choice was the rational choice for at least 1 participant.}\label{tab:excleader}
\begin{tabular}{lcccc}
\toprule
Number of participants with A & None & Individual-level & Group-level & Total \\
 as their only rational choice & & LLM facilitation & LLM facilitation & \\
\midrule
0   & 12.5 (1/8)   & 50.0 (3/6)   & 60.0 (3/5)   & 36.8 (7/19)  \\
1   & 25.0 (5/20)  & 44.4 (12/27) & 38.9 (7/18)  & 36.9 (24/65) \\
2   & 0.0 (0/3)    & 100.0 (1/1)  & 75.0 (3/4)   & 50.0 (4/8)   \\
3   & 60.0 (9/15)  & 83.3 (10/12) & 68.4 (13/19) & 69.6 (32/46) \\
\midrule
All & 32.6 (15/46) & 56.5 (26/46) & 56.5 (26/46) & 48.6 (67/138) \\
\botrule
\end{tabular}
\label{sitab:rational}
\end{table}
\clearpage
 \begin{table}[ht]
\centering
\caption{Participant demographics and self-rated familiarity with AI ($N = 414$). The AI familiarity question was phrased as: ``How much knowledge do you have in artificial intelligence?'' based on \cite{yougov2024ai}.}
\label{sitab:participant-demographics}
\begin{tabular}{llrr}
\toprule
\textbf{Characteristic} & \textbf{Category} & \textbf{$n$} & \textbf{\%} \\
\midrule

\textbf{Age}
    & 18--24       & 45  & 10.9 \\
    & 25--34       & 122 & 29.5 \\
    & 35--44       & 115 & 27.8 \\
    & 45--54       & 89  & 21.5 \\
    & Over 55      & 43  & 10.4 \\
\addlinespace

\textbf{Gender}
    & Male                 & 216 & 52.2 \\
    & Female               & 192 & 46.4 \\
    & Prefer not to say    & 6   & 1.4 \\
\addlinespace

\textbf{Education}
    & Some high school     & 3   & 0.7 \\
    & High school          & 124 & 30.0 \\
    & Bachelor's degree    & 199 & 48.1 \\
    & Master's degree      & 72  & 17.4 \\
    & PhD or higher        & 13  & 3.1 \\
    & Prefer not to say    & 3   & 0.7 \\
\addlinespace

\textbf{Ethnicity}
    & Caucasian             & 266 & 64.3 \\
    & Asian                 & 51  & 12.3 \\
    & African American      & 50  & 12.1 \\
    & Latino or Hispanic    & 20  & 4.8 \\
    & Two or more ethnicities & 17 & 4.1 \\
    & Native American       & 4   & 1.0 \\
    & Other                 & 2   & 0.5 \\
    & Prefer not to say     & 4   & 1.0 \\
\addlinespace

\textbf{AI familiarity}
    & None                  & 2   & 0.5 \\
    & Low                   & 34  & 8.2 \\
    & Average               & 207 & 50.0 \\
    & Moderately high       & 149 & 36.0 \\
    & High                  & 22  & 5.3 \\

\bottomrule
\end{tabular}
\end{table}

\end{document}